\documentclass[11pt,twoside]{article}
\usepackage{mathrsfs}
\usepackage{amsmath}
\usepackage{amssymb}
\usepackage{graphicx}
\usepackage{subfigure}
\usepackage{appendix}
\usepackage{cite}
\numberwithin{equation}{section}

\newtheorem{proposition}{Proposition}

\DeclareMathOperator{\im}{Im}
\DeclareMathOperator{\re}{Re}
\DeclareMathOperator{\tr}{tr}
\newcommand*{\QEDB}{\hfill\ensuremath{\square}}

\title{On the nonlinear Schr\"{o}dinger equation with a time-dependent boundary condition}
\author{
Baoqiang Xia
\\
School of Mathematics and Statistics, Jiangsu Normal University,
\\
Xuzhou, Jiangsu 221116, P. R. China,
\\
E-mail address: xiabaoqiang@126.com
}

\date{}
\begin{document}
\maketitle
\begin{abstract}
We study the nonlinear Schr\"{o}dinger equation on the half-line with a boundary condition that involves time derivative.
This boundary condition was presented by Zambon [J. High Energ. Phys. 2014 (2014) 36].
We establish the integrability of such a boundary both by using the Sklyanin's formalism and by using the tool of B\"{a}cklund transformations together with a suitable reduction of reflection type.
Moreover, we present a method to derive explicit formulae for multi-soliton solutions of the boundary problem by virtue of the Darboux transformation method in conjunction with a boundary dressing technique.

\noindent {\bf Keywords:}\quad integrable boundary conditions, B\"{a}cklund transformation, Darboux transformation, soliton solutions.

\end{abstract}
\newpage

\section{ Introduction}

We study the nonlinear Schr\"{o}dinger (NLS) equation
\begin{eqnarray}
iu_t+u_{xx}+2 u|u|^2=0,
\label{NLS}
\end{eqnarray}
posed on the positive $x$-axis with the following boundary condition \cite{Zambon2014}
\begin{eqnarray}
\left.\left(iu_t-2u_{x}\sqrt{4b^2-|u|^2}-4(a^2+b^2)u+2 u|u|^2\right)\right|_{x=0}=0,
\label{tbc}
\end{eqnarray}
where $a$ and $b$ are two arbitrary real constants.

The boundary condition (\ref{tbc}) was presented by Zambon in  \cite{Zambon2014} via dressing a Dirichlet boundary with an integrable defect of type I for the NLS equation.
In this paper we will not concentrate on the subject of integrable defect systems. We refer the reader to \cite{BCZ20041,CZ2006,HK2008,Caudrelier2008,CK2015,AD2012,Doikou2016,CP2017} and references therein on this interesting subject.
Due to the presence of the time derivative in the boundary term, we will refer to (\ref{tbc}) as a time-dependent boundary condition in this paper.
It is worth mentioning that an analogous boundary for an integrable discrete model, the Ablowitz-Ladik system, was presented and investigated very recently in \cite{CC2019}.

The integrability of the boundary condition (\ref{tbc}) has been established in \cite{Zambon2014} by using the modified Lax pair technique in conjunction with the classical $r$-matrix method (see e.g. \cite{Faddeev2007}).
It was shown that the boundary matrix (the solution of the reflection equation) associated with (\ref{tbc}), unlike the constant one found in the Sklyanin's formalism \cite{Sklyanin1987}, is dynamical: it depends on the field of the NLS equation at the boundary location.
In other words, one needs to consider the dynamical generalization of the Sklyanin's formalism in order to study the integrability of the boundary condition (\ref{tbc}) via $r$-matrix method.

One of the aims of the present paper is to describe a method, which is different from the one presented in \cite{Zambon2014},
for investigating the integrability of the time-dependent boundary (\ref{tbc}).
This method is based on B\"{a}cklund transformations (BTs) together with a suitable reduction of reflection type.
It is worth reminding that the idea to study boundary conditions compatible with integrability via BTs was initiated by Habibullin in \cite{Habibullin1991},
where the analysis is based on the space-part of BTs.
The main difference with respect to the method in \cite{Habibullin1991} is that our argument is based on the time-part of the BTs.
We first present a connection between the Sklyanin's formalism and the time-part of the BTs.
Based on this observation we are able to construct time-dependent boundary conditions directly from BTs combined with a reduction, without dressing a known integrable boundary.
The advantage of the method is that it enables us to construct a generating function of the modified conserved quantities to support the integrability of the boundary (\ref{tbc}) directly from the Lax pair formulation.

Another aim of the present paper is to study soliton solutions of the NLS equation (\ref{NLS}) together with the boundary condition (\ref{tbc}).
We present an approach for constructing soliton solutions for such a time-dependent boundary problem, which is based on the tool of Darboux transformations (DTs) \cite{MS1991} in conjunction with a boundary dressing technique \cite{Zhang2019}.
As a consequence, we obtain explicit formulae for multi-soliton solutions of the NLS equation in the presence of the boundary condition (\ref{tbc}).

The paper is organized as follows.
In section 2, we establish the integrability of the boundary condition (\ref{tbc}) both by using the Sklyanin's formalism and by using the tool of B\"{a}cklund transformations together with a reduction technique. In section 3, we show how to calculate soliton solutions of the NLS equation with the integrable boundary condition (\ref{tbc}).
Some concluding remarks are drawn in section 4.

\section{Integrability of the time-dependent boundary condition}

For self-containedness, we start by a brief summary of the constructions of conservation laws and the BTs for the NLS equation.
The NLS equation (\ref{NLS}) in the bulk admits the following Lax pair formulation
\begin{subequations}
\begin{eqnarray}
\phi_x(x,t,\lambda)=U(x,t,\lambda)\phi(x,t,\lambda),
 \label{lpx}
 \\
\phi_t(x,t,\lambda)=V(x,t,\lambda)\phi(x,t,\lambda),
 \label{lpt}
\end{eqnarray}
\label{LPxt}
\end{subequations}
where $\lambda$ is a spectral parameter, $\phi=(\phi_{1},~\phi_{2})^T$, and
\begin{eqnarray}
U=\left( \begin{array}{cc} -i\lambda & u \\
 -\bar{u} &  i\lambda \\ \end{array} \right),
~~ V=\left( \begin{array}{cc}  -2i\lambda^2+i|u|^2 & 2\lambda u+iu_x \\
 -2\lambda \bar{u}+i\bar{u}_x & 2i\lambda^2-i|u|^2  \\ \end{array} \right).
 \label{UV}
\end{eqnarray}
Here and in what follows the bar indicates complex conjugation.
By using the Lax pair formulation, one can construct an infinite set of conservation laws for the NLS equation with the condition of vanishing boundary or periodic boundary.
Indeed, denoting $\Gamma=\frac{\phi_2}{\phi_1}$, we find from (\ref{LPxt}) the following $x$-part and $t$-part Riccati equations
\begin{subequations}
\begin{eqnarray}
\Gamma_x=2i\lambda \Gamma-\bar{u}-u\Gamma^2,
\label{ricx}
\\
\Gamma_t=-2\lambda \bar{u}+i\bar{u}_x -2\left(-2i\lambda^2+i|u|^2\right) \Gamma-\left(2\lambda u+iu_x \right)\Gamma^2,
\label{rict}
\end{eqnarray}
\label{ric}
\end{subequations}
together with the following conservation equation
\begin{eqnarray}
\left(u\Gamma\right)_t=\left(i|u|^2+\left(2\lambda u+iu_x\right)\Gamma\right)_x.
\label{CL}
\end{eqnarray}
Using the vanishing boundary condition or the periodic boundary condition, equation (\ref{CL}) implies that the function $u\Gamma$ provides a generating function of the conservation densities.
By substituting the expansion
$\Gamma=\sum_{n=1}^{\infty}\Gamma_n(2i\lambda)^{-n}$
into (\ref{ricx}) and by equating the coefficients of powers of $\lambda$, we find explicit forms of conservation
densities $u\Gamma_n$ with
\begin{eqnarray}
\Gamma_{1}=\bar{u}, ~~\Gamma_{2}=\bar{u}_x,
~~\Gamma_{n+1}=\left(\Gamma_{n}\right)_x+u\sum_{j=1}^{n-1}\Gamma_j\Gamma_{n-j},\quad n\geq 2.
\label{wj}
\end{eqnarray}

In order to study BTs for the NLS equation, we introduce another copy of the auxiliary problems for $\tilde{\phi}$ with Lax pair
$\tilde{U}$, $\tilde{V}$ defined as in (\ref{UV}) with the new potentials $\tilde{u}$, $\tilde{v}$, replacing $u$, $v$.
We assume that the two systems are related by the gauge transformation,
\begin{eqnarray}
\phi(x,t,\lambda)=B(x,t,\lambda)\tilde{\phi}(x,t,\lambda),
\label{BT}
\end{eqnarray}
where the matrix $B(x,t,\lambda)$ satisfies
\begin{subequations}
\begin{eqnarray}
B_x(x,t,\lambda)=U(x,t,\lambda)B(x,t,\lambda)-B(x,t,\lambda)\tilde{U}(x,t,\lambda),
\label{BT1a}
\\
B_t(x,t,\lambda)=V(x,t,\lambda)B(x,t,\lambda)-B(x,t,\lambda)\tilde{V}(x,t,\lambda).
\label{BT1b}
\end{eqnarray}
\label{BT1}
\end{subequations}
For the NLS equation (\ref{NLS}), we take
\begin{eqnarray}
B=\left(\lambda-a\right) I+\frac{1}{2}\left( \begin{array}{cc} -i\Omega & i(\tilde{u}-u)\\
 i (\tilde{u}-u)^* &  i\Omega  \\ \end{array} \right),
~~\Omega= \sqrt{4b^2 - |u-\tilde{u}|^2},
 \label{nlsdm}
\end{eqnarray}
Then equation (\ref{BT1}) induces the following BT between the potentials $u$ and $\tilde{u}$:
\begin{subequations}
\begin{eqnarray}
u_x-\tilde{u}_x=-2ia\left(u-\tilde{u}\right)-\left(u+\tilde{u}\right)\Omega,
\label{BTNLSa}
\\
u_t-\tilde{u}_t=2a\left(u_x-\tilde{u}_x\right)-i\Omega \left(\tilde{u}_x+u_x\right)+i(u-\tilde{u})\left(|u|^2+|\tilde{u}|^2\right).
\label{BTNLSb}
\end{eqnarray}
\label{BTNLS}
\end{subequations}

Following \cite{Zambon2014}, we now show how the boundary condition (\ref{tbc}) can be derived.
The derivation is based on dressing a Dirichlet boundary with an integrable defect condition of the NLS equation.
In order to describe the defect conditions, we denote by $\tilde{u}$ the field on the negative $x$-axis and by $u$ the field on the positive $x$-axis.
For the NLS equation, the defect conditions of type I are defined by the BT (\ref{BTNLS}) frozen at the defect location $x=0$ rather than on the whole line (see, for example \cite{CZ2006,Caudrelier2008}, for details).
We assume that the field $\tilde{u}$ at the boundary $x=0$ satisfies the Dirichlet boundary condition,
that is $\left.\tilde{u}\right|_{x=0}=0$ and $\left.\tilde{u}_t\right|_{x=0}=0$.
In this situation, the corresponding defect conditions become
\begin{subequations}
\begin{eqnarray}
\left.\left(u_x-\tilde{u}_x\right)\right|_{x=0}=\left.\left(-2iau-u\sqrt{4b^2 - |u|^2}\right)\right|_{x=0},
\label{BTNLSr1}
\\
\left.u_t\right|_{x=0}=\left.\left(2a\left(u_x-\tilde{u}_x\right)-i\left(\tilde{u}_x+u_x\right)\sqrt{4b^2 - |u|^2}+iu|u|^2\right)\right|_{x=0},
\label{BTNLSr2}
\end{eqnarray}
\label{BTNLSr}
\end{subequations}
By using (\ref{BTNLSr1}) to eliminate $\tilde{u}_x$ in (\ref{BTNLSr2}) we obtain the time-dependent boundary condition (\ref{tbc}).

\subsection{Interpretation of integrability of the boundary via the classical $r$-matrix}\label{sec2.2}

We now discuss the integrability of the time-dependent boundary (\ref{tbc}) via the classical $r$-matrix method.
The analysis is based on an extension of the boundary $K(\lambda)$ matrix in Sklyanin's formalism from non-dynamical case to a dynamical (time-dependent) case.
We note that most of the arguments in this subsection are essentially the same as those in \cite{Zambon2014}
(see also \cite{ACC2018} for recent studies on the dynamical consideration of Sklyanin's formalism).
The new results are that we find a general solution of the dynamical reflection equation (see (\ref{K}) below),
and present a canonical realization of the associated boundary Poisson brackets (see (\ref{S123qp}) below).

We first sketch out the standard Sklyanin's formalism for studying integrable boundary conditions.
For the NLS equation, we consider the canonical Poisson brackets
\begin{eqnarray}
\left\{u(x,t),u(y,t)\right\}=\left\{\bar{u}(x,t),\bar{u}(y,t)\right\}=0,
~~
\left\{u(x,t),\bar{u}(y,t)\right\}=i\delta(x-y),
\label{cPB}
\end{eqnarray}
where $\delta(x-y)$ is the Dirac $\delta$-function.
It follows that the transition matrix,
\begin{eqnarray}
T(x,y,\lambda)=\overset{\curvearrowleft}\exp \int_{y}^{x} U(\xi,t,\lambda)d\xi,
\end{eqnarray}
satisfies the following well-known relation (see e.g. \cite{Faddeev2007})
\begin{eqnarray}
\left\{T_1(x,y,\lambda),T_2(x,y,\mu)\right\}=\left[r(\lambda-\mu),T_1(x,y,\lambda)T_2(x,y,\mu)\right],
\label{rmrelation}
\end{eqnarray}
where $T_1(x,y,\lambda)=T(x,y,\lambda)\otimes I$, $T_2(x,y,\mu)=I\otimes T(x,y,\mu)$, and the classical $r$-matrix $r$ is
\begin{eqnarray}
r(\lambda)=\frac{1}{2\lambda}\left( \begin{array}{cccc}
1 & 0 & 0 & 0
\\
0 &  0 & 1 & 0
\\
0 & 1 & 0 & 0
\\
0 &  0 & 0 & 1
 \\ \end{array} \right).
 \label{rm}
\end{eqnarray}
In the study of integrable boundary conditions on the half-line,
Sklyanin in his seminal paper \cite{Sklyanin1987} presented a generalization of the monodromy matrix $T(\lambda)$, which is
\begin{eqnarray}
\tau(\lambda)=T(\lambda)K(\lambda)T^{-1}(-\lambda),
\label{gm}
\end{eqnarray}
where $T(\lambda)=T(\infty,0,\lambda)$.
In the case of $K(\lambda)$ being a constant matrix, it can, by using (\ref{rmrelation}), be shown that if $K(\lambda)$ satisfies the relation (called reflection equation)
\begin{eqnarray}
\left[r(\lambda-\mu),K_1(\lambda)K_2(\mu)\right]+K_1(\lambda)r(\lambda+\mu)K_2(\mu)-K_2(\mu)r(\lambda+\mu)K_1(\lambda)=0,
\end{eqnarray}
then the quantities $\tr(\tau(\lambda))$ are in involution for different values of the spectral parameter:
\begin{eqnarray}
\left\{\tr(\tau(\lambda)),\tr(\tau(\mu))\right\}=0.
\end{eqnarray}
Moreover, if the boundary satisfies the condition
\begin{eqnarray}
V(0,t,\lambda)K(\lambda)=K(\lambda)V(0,t,-\lambda),
\end{eqnarray}
then $\frac{d\tr(\tau(\lambda))}{dt}=0$.
Thus $\tr(\tau(\lambda))$ provides a generating function of Poisson commuting integrals of motion.
The well-known Dirichlet boundary,  Neumann boundary and Robin boundary conditions and their integrability can be concluded by choosing the matrix $K(\lambda)$ that is proportional to some constant diagonal matrices (see \cite{Sklyanin1987} for details).

In order to study the integrability of the time-dependent boundary, we need to consider the dynamical generalisation of the $K(\lambda)$ matrix.
In this case, we can deduce that if $K(\lambda)$ satisfies relations
\begin{eqnarray}
\left\{K_1(\lambda),K_2(\mu)\right\}=\left[r(\lambda-\mu),K_1(\lambda)K_2(\mu)\right]+K_1(\lambda)r(\lambda+\mu)K_2(\mu)-K_2(\mu)r(\lambda+\mu)K_1(\lambda),
 \label{alga}
 \\
\left\{K_1(\lambda),U_2(x,t,\mu)\right\} =0,
 \label{algb}
\end{eqnarray}
 then the quantities $\tr(\tau(\lambda))$ are in involution: $\left\{\tr(\tau(\lambda)),\tr(\tau(\mu))\right\}=0$.
 Moreover, if the $K(\lambda)$ matrix satisfies the equation
\begin{eqnarray}
 \frac{dK(\lambda)}{dt}=V(0,t,\lambda)K(\lambda)-K(\lambda)V(0,t,-\lambda),
 \label{algc}
\end{eqnarray}
 then $\frac{d\tr(\tau(\lambda))}{dt}=0$.
 We will restrict our attention to the case that the $K(\lambda)$ matrix holds at the boundary location.
 In this case, the Poisson bracket (\ref{algb}) is automatically zero.
 For the dynamical reflection equation (\ref{alga}), we find the following solution
\begin{eqnarray}
K(\lambda)=\left( \begin{array}{cc} \lambda-c\lambda^{-1} & 0 \\
 0 &  \lambda-c\lambda^{-1} \\ \end{array} \right)+\left( \begin{array}{cc} S_3 & S_1 \\
 S_2 &  -S_3 \\ \end{array} \right),
 \label{K}
\end{eqnarray}
where $c$ is an arbitrary constant, and the dynamical variables $\left(S_1,S_2,S_3\right)$ obey to the following Poisson brackets:
\begin{eqnarray}
\left\{S_3,S_1\right\}=S_1,~~\left\{S_3,S_2\right\}=-S_2,~~\left\{S_1,S_2\right\}=2S_3.
\label{sl2PB}
\end{eqnarray}
Inserting (\ref{K}) into (\ref{algc}), we find
\begin{eqnarray}
\begin{split}
S_1=\left.-iu\right|_{x=0},~~
S_2=\left.-i\bar{u}\right|_{x=0},
~~
S_3=\left.-i\sqrt{\alpha^2-|u|^2}\right|_{x=0},
\end{split}
\label{S123}
\end{eqnarray}
and the following boundary condition
 \begin{eqnarray}
\left.\left(i\frac{du}{dt}-4cu+2 u|u|^2-2u_{x}\sqrt{\alpha^2-|u|^2}\right)\right|_{x=0}=0,
\label{gbc}
\end{eqnarray}
where $c$ and $\alpha$ are arbitrary real constants.
By setting $\alpha^2=4b^2$ and $c=a^2+b^2$ in (\ref{gbc}), we recover the boundary condition (\ref{tbc}).
This implies the integrability of the boundary condition (\ref{tbc}).

As mentioned above, the boundary quantities $\left(S_1,S_2,S_3\right)$, given by (\ref{S123}), should obey to the Poisson brackets (\ref{sl2PB}), in order to guarantee the Poisson commutativity of integrals of motion associated with the boundary condition (\ref{gbc}).
Next we show $\left(S_1,S_2,S_3\right)$ admit a natural canonical realization.
Indeed, we rewrite the field of the NLS equation as
\begin{eqnarray}
u(x,t)=\frac{i}{2}p^2(x,t),
~~
\bar{u}(x,t)=-\frac{i}{2}q^2(x,t)-2i\alpha^2p^{-2}(x,t),
\label{fcanonb}
\end{eqnarray}
and denote the boundaries of $p$ and $q$ at $x=0$ by
\begin{eqnarray}
p_0(t)=p(0,t),
~~
q_0(t)=q(0,t),
\label{pq}
\end{eqnarray}
In this case, $\left(S_1,S_2,S_3\right)$, defined by (\ref{S123}), become
\begin{eqnarray}
\begin{split}
S_1=\frac{1}{2}p_0^2(t),~~
S_2=-\frac{1}{2}q_0^2(t)-2\alpha^2p_0^{-2}(t),
~~
S_3=\frac{1}{2}q_0(t)p_0(t).
\end{split}
\label{S123qp}
\end{eqnarray}
One can check directly that $\left(S_1,S_2,S_3\right)$, given by (\ref{S123qp}),
satisfy the algebra (\ref{sl2PB}), if the variables $q_0(t)$ and $p_0(t)$ satisfy the canonical Poisson brackets, that is
\begin{eqnarray}
\left\{q_0(t),q_0(t)\right\}=\left\{p_0(t),p_0(t)\right\}=0,
~~
\left\{q_0(t),p_0(t)\right\}=1.
\label{cPB1}
\end{eqnarray}
Thus (\ref{S123qp}) does provide a canonical realization of the boundary algebra (\ref{sl2PB}).

\subsection{Interpretation of integrability of the boundary as a direct reduction of BTs}

In this subsection, we will describe a method for studying the time-dependent boundary conditions,
which is different from the one presented above and which is based on the BTs.

We first describe how the time-part of BTs is related to the Sklyanin's formalism.
The key point is the following observation: if there exists a compatible reduction on the fields $u$ and $\tilde{u}$ such that
\begin{eqnarray}
\tilde{V}(0,t,\lambda)=V(0,t,-\lambda),
\label{V0}
\end{eqnarray}
then the time-part of BT (\ref{BT1b}) evaluated at $x=0$ is equivalent to the boundary equation (\ref{algc}) appearing in Sklyanin's formalism in the dynamical case.
The odd reduction $\tilde{u}(x,t)=-u(-x,t)$ meets perfectly this requirement for the reduction on fields.
Indeed, one can check directly that the NLS equation admits the odd reduction $\tilde{u}(x,t)=-u(-x,t)$ and equation (\ref{V0}) becomes an identity under this reduction.
As a consequence, the time-part of BT (\ref{BT1b}) with the reduction $\tilde{u}(x,t)=-u(-x,t)$ evaluated at $x=0$ is equivalent to the boundary equation (\ref{algc}).

The above analysis implies that we can derive the time-dependent boundary condition (\ref{tbc}) directly via imposing the the odd reduction $\tilde{u}(x,t)=-u(-x,t)$ on two fields $u$ and $\tilde{u}$ related by BT (\ref{BTNLS}). Indeed, by inserting (\ref{BTNLSa}) into (\ref{BTNLSb}) we obtain
\begin{eqnarray}
u_t-\tilde{u}_t=-4ia^2\left(u-\tilde{u}\right)-\Omega\left(2a\left(u+\tilde{u}\right)+i \left(\tilde{u}_x+u_x\right)\right)+i(u-\tilde{u})\left(|u|^2+|\tilde{u}|^2\right).
\label{BTNLSc}
\end{eqnarray}
By imposing the reduction $\tilde{u}(x,t)=-u(-x,t)$ on (\ref{BTNLSc}) and evaluating the resulting equation at $x=0$,
we then recover the boundary condition (\ref{tbc}) up to a slight scaling.

In general, it is not easy to find a dynamical boundary $K$ matrix that matches the boundary equation (\ref{algc}), a guess work is usually employed.
The analysis presented above provides us a hint to deduce such a boundary matrix, that is the $K$ matrix can be derived from BTs together with a reduction of reflection type.
Indeed, by inserting the reduction $\tilde{u}(x,t)=-u(-x,t)$ into the B\"{a}cklund matrix (\ref{nlsdm}) evaluated at $x=0$
and after a slight adjustment to the diagonal term of the resulting matrix,
we recover the boundary $K$ matrix defined by (\ref{K}) and (\ref{S123}) in section \ref{sec2.2}.

Another advantage of the method is that we can construct explicitly the generating function of the infinite set of modified conservation laws to support the integrability of the boundary directly from the Lax pair formulation. The result is as follows.
\begin{proposition}\label{pro1}
A generating function for the integrals of motion of the NLS with integrable boundary condition (\ref{tbc}) is given by
\begin{eqnarray}
\begin{split}
I(\lambda)=&\int_{0}^{\infty}u(x,t)\left(\Gamma (x,t,\lambda)-\Gamma (x,t,-\lambda)\right)dx
\\&+\left.\ln\left(\lambda-(a^2+b^2)\lambda^{-1}-i\sqrt{4b^2-|u(x,t)|^2}-iu(x,t)\Gamma (-x,t,-\lambda)\right)\right|_{x=0},
\end{split}
\label{GCD}
\end{eqnarray}
where $\Gamma(x,t,\lambda)$ satisfies the Ricatti equation (\ref{ricx}) (see (\ref{wj}) for explicit forms of $\Gamma$).
\end{proposition}
{\bf Proof} Consider two copies of the auxiliary problem (\ref{LPxt}) related by (\ref{BT}).
Following \cite{Caudrelier2008}, we have
\begin{eqnarray}
\frac{d}{dt}\left(\int_{-\infty}^{0}\tilde{u}\tilde{\Gamma}(x,t,\lambda) dx+\int_{0}^{\infty}u\Gamma (x,t,\lambda)dx
+\ln\left(B_{11}(0,t,\lambda)+B_{12}\Gamma(0,t,\lambda)\right)\right)=0,
\label{IM}
\end{eqnarray}
where $B_{jk}$, $j,k=1,2$, is the $jk$-entry of the defect matrix $B$.
For the boundary problem (\ref{tbc}), under the reduction $\tilde{u}(x,t)=-u(-x,t)$ we have
\begin{eqnarray}
\tilde{\Gamma}(x,t,\lambda)=\Gamma(-x,t,-\lambda).
\label{redGram}
\end{eqnarray}
Using (\ref{redGram}) in  (\ref{IM}), we obtain the integrals of motion (\ref{GCD}) after some algebra.
\QEDB

By inserting (\ref{wj}) into (\ref{GCD}), we immediately obtain explicit forms for the conservation densities.
For example, the first few members are
\begin{eqnarray}
\begin{split}
I_1=&-i\int_0^{\infty}|u|^2dx-\left.i\Omega(u)\right|_{x=0},
\\
I_3=&\frac{i}{4}\int_0^{\infty}\left(|u|^4-|u_x|^2\right)dx+\left.\left[\frac{i}{6}|u|^2-i(a^2-\frac{b^2}{3})\right]\Omega(u)\right|_{x=0},
\\
I_5=&-\frac{i}{16}\int_0^{\infty}u\left(\bar{u}_{xxxx}+6\bar{u}|u_x|^2+\bar{u}^2u_{xx}+5u(\bar{u}_x)^2+|u|^2(6\bar{u}_{xx}+2\bar{u}|u|^2)\right)dx
\\&-\left.\left[\frac{i}{16}u\bar{u}_{xxx}+\frac{i}{16}|u|^2\bar{u}u_x+\frac{3i}{8}|u|^2u\bar{u}_x-\frac{i}{4}(a^2+b^2)u\bar{u}_x\right]\right|_{x=0}
\\&
-\left.i\Omega(u)\left(\frac{1}{8}u\bar{u}_{xx}+\frac{3}{8}|u|^4+(a^2+b^2)(a^2+b^2-|u|^2)\right)\right|_{x=0}
\\&
-\left[\left.\frac{i}{4}u\bar{u}_x(\Omega(u))^2-i(a^2+b^2-\frac{1}{2}|u|^2)(\Omega(u))^3+\frac{i}{5}(\Omega(u))^5\right]\right|_{x=0},
\end{split}
\label{ecd}
\end{eqnarray}
where $\Omega(u)=\sqrt{4b^2-|u(x,t)|^2}$.
These formulae coincide with the ones found in \cite{Zambon2014}.

\section{Soliton solutions meeting the boundary condition}

In this section we will show how to derive soliton solutions of the NLS equation in the presence of the boundary condition (\ref{tbc}).
Our method is based on the tool of DTs \cite{MS1991} in conjunction with a boundary dressing technique \cite{Zhang2019}.

The well-known DT for the NLS equation is constructed as follows \cite{MS1991}. If $\tilde{u}$ and $(\tilde{\phi}_{1},\tilde{\phi}_{2})^T$ satisfy differential equations (\ref{LPxt}), then so does
\begin{subequations}
\begin{eqnarray}
u=\tilde{u}-2i(\xi_1-\bar{\xi}_1)\frac{f_{1}\bar{f}_{2}}{|f_{1}|^2+|f_{2}|^2},
\label{DTa}
\\
(\phi_{1},\phi_{2})^T=D(x,t,\lambda)(\tilde{\phi}_{1},\tilde{\phi}_{2})^T,
\end{eqnarray}
\label{DT}
\end{subequations}
where $(f_{1},f_{2})^T$ is a special solution of the linear auxiliary system (\ref{LPxt}) with $\tilde{u}$ and with $\lambda=\xi_1$,
and
\begin{eqnarray*}
D(x,t,\lambda)=(\lambda-\bar{\xi}_1)I+\frac{\bar{\xi}_{1}-\xi_1}{|f_{1}|^2+|f_{2}|^2}\left( \begin{array}{cc} |f_{1}|^2 & f_1\bar{f}_2 \\
 \bar{f}_1f_2 &  |f_{2}|^2 \\ \end{array} \right).
\end{eqnarray*}

We now show that $u$ and $\tilde{u}$ connected by (\ref{DTa}) also satisfy a BT of the same form as (\ref{BTNLS}).
Indeed, differentiating (\ref{DT}) with respect to $x$ and using
$$
f_{1,x}=-i\xi_1f_{1}+ \tilde{u}f_{2}, ~~f_{2,x}=-\bar{\tilde{u}}f_{1}+i\xi_1f_{2},
$$
we obtain
\begin{eqnarray}
u_{x}-\tilde{u}_{x}=-2i\re\xi_1 \left(u-\tilde{u}\right)-\left(u+\tilde{u}\right)\sqrt{4\left(\im\xi_1\right)^2 - |u-\tilde{u}|^2}.
\label{DTd}
\end{eqnarray}
Setting $\xi_1=a+ib$, equation (\ref{DTd}) becomes (\ref{BTNLSa}).
Similarly, by differentiating (\ref{DT}) with respect to $t$ and by using the fact that $(f_{1},f_{2})^T$ satisfies (\ref{lpt}) with $\tilde{u}$ and $\lambda=\xi_1$,
we obtain (\ref{BTNLSb}).

In light of the above arguments,
we can use the DT to construct the solution of the NLS equation in the presence of the boundary condition (\ref{tbc}).
The strategy consists of the following two steps:
(1) find a solution $\tilde{u}$ of the NLS equation satisfying the Dirichlet boundary condition $\tilde{u}|_{x=0}=0$ and find a solution $(f_{1},f_{2})^T$ of the corresponding Lax pair system;
(2) by using the DT (\ref{DT}) with $\xi_1=a+ib$, construct a new solution $u$ of the NLS equation.
Such a solution $u$ satisfies the time-dependent boundary condition (\ref{tbc}), due to the fact that the boundary (\ref{tbc}) can be derived by dressing the Dirichlet boundary with the integrable defect condition (the frozen BT).
In particular, if we find a $N$-soliton solution of $\tilde{u}$ in the first step,
then we can, after using the second step, generate the $(N+1)$-soliton solution $u$ satisfying the boundary condition (\ref{tbc}).

We next implement the above two steps in details.
It is evident that the trivial solution $\tilde{u}=0$ satisfies the Dirichlet boundary condition.
The solution $(f_{1}(\xi_1),f_{2}(\xi_1))^T$ of the associated Lax pair system can be taken in the form of
\begin{eqnarray}
(f_{1}(\xi_1),f_{2}(\xi_1))^T=\left(A_1e^{-i\xi_1x-2i\xi_1^2t},B_1e^{i\xi_1x+2i\xi_1^2t}\right)^T,
\label{0phi}
\end{eqnarray}
where $\xi_1=a+ib$, and $A_1$ and $B_1$ are arbitrary constants.
By substituting (\ref{0phi}) into (\ref{DT}), we obtain the following single soliton satisfying the boundary condition (\ref{tbc}),
\begin{eqnarray}
u=4b\frac{f_{1}\bar{f}_{2}}{|f_{1}|^2+|f_{2}|^2}=2b\frac{\exp(-2iax-4i(a^2-b^2)t+i\theta_1)}{\cosh(bx+4abt+\theta_2)},
\label{1stbc}
\end{eqnarray}
where $\exp(i\theta_1)=\frac{A_1\bar{B}_1}{|A_1||B_1|}$ and $\exp(\theta_2)=\left|\frac{A_1}{B_1}\right|$.

It has been shown that non-trivial soliton solutions $\tilde{u}$ satisfying the Dirichlet boundary condition can be constructed by using the DT \cite{Zhang2019}.
More precisely, the $N$-soliton solution, by applying the DT to dress the Dirichlet boundary, are given by
\begin{eqnarray}
\tilde{u}[n]=2i\frac{\Delta_2[N]}{\Delta_1[N]},
\label{ns}
\end{eqnarray}
where
\begin{eqnarray}
\Delta_1[N]=\left| \begin{array}{cccccccc} \lambda_1^{2N-1}\nu_1 & \lambda_2^{2N-1}\nu_2 & \cdots &  \lambda_{2N}^{2N-1}\nu_{2N}
& \bar{\lambda}_1^{2N-1}\bar{\mu}_1 & \bar{\lambda}_2^{2N-1}\bar{\mu}_2 & \cdots &  \bar{\lambda}_{2N}^{2N-1}\bar{\mu}_{2N} \\
\lambda_1^{2N-2}\nu_1 & \lambda_2^{2N-2}\nu_2 & \cdots &  \lambda_{2N}^{2N-2}\nu_{2N}
& \bar{\lambda}_1^{2N-2}\bar{\mu}_1 & \bar{\lambda}_2^{2N-2}\bar{\mu}_2 & \cdots &  \bar{\lambda}_{2N}^{2N-2}\bar{\mu}_{2N} \\
\cdots & \cdots & \cdots & \cdots & \cdots & \cdots & \cdots & \cdots \\
\nu_1 & \nu_2 & \cdots &  \nu_{2N}
& \bar{\mu}_1 & \bar{\mu}_2 & \cdots &  \bar{\mu}_{2N}\\
\lambda_1^{2N-1}\mu_1 & \lambda_2^{2N-1}\mu_2 & \cdots &  \lambda_{2N}^{2N-1}\mu_{2N}
& -\bar{\lambda}_1^{2N-1}\bar{\nu}_1 & -\bar{\lambda}_2^{2N-1}\bar{\nu}_2 & \cdots &  -\bar{\lambda}_{2N}^{2N-1}\bar{\nu}_{2N} \\
\lambda_1^{2N-2}\mu_1 & \lambda_2^{2N-2}\mu_2 & \cdots &  \lambda_{2N}^{2N-2}\mu_{2N}
& -\bar{\lambda}_1^{2N-2}\bar{\nu}_1 & -\bar{\lambda}_2^{2N-2}\bar{\nu}_2 & \cdots &  -\bar{\lambda}_{2N}^{2N-2}\bar{\nu}_{2N} \\
\cdots & \cdots & \cdots & \cdots & \cdots & \cdots & \cdots & \cdots \\
\mu_1 & \mu_2 & \cdots &  \mu_{2N}
& -\bar{\nu}_1 & -\bar{\nu}_2 & \cdots &  -\bar{\nu}_{2N}\\
\end{array} \right|,
 \label{delta1n}
 \end{eqnarray}
and $\Delta_2[N]$ are defined by the same determinant as $\Delta_1[N]$ but with the first row replaced by
$$\left(-\lambda_1^{2N}\mu_1, -\lambda_2^{2N}\mu_2, \cdots, -\lambda_{2N}^{2N}\mu_{2N},
\bar{\lambda}_1^{2N}\bar{\nu}_1, \bar{\lambda}_2^{2N}\bar{\nu}_2, \cdots,  \bar{\lambda}_{2N}^{2N}\bar{\nu}_{2N}\right),$$
and
\begin{eqnarray}
\begin{split}
\lambda_{2j}=-\lambda_{2j-1}, ~~\bar{\lambda}_{2j-1}\neq -\lambda_{2j-1},~~\lambda_{2k-1}\neq -\lambda_{2j-1},~~j=1,\cdots, N,
\\
\mu_{2j-1}=c_{j}e^{-i\lambda_{2j-1}x-2i\lambda_{2j-1}^2t}, ~~\nu_{2j-1}=d_{j}e^{i\lambda_{2j-1}x+2i\lambda_{2j-1}^2t}, ~~j=1,\cdots, N,
\\
\mu_{2j}=c_{j}e^{i\lambda_{2j-1}x-2i\lambda_{2j-1}^2t}, ~~\nu_{2j}=d_{j}e^{-i\lambda_{2j-1}x+2i\lambda_{2j-1}^2t}, ~~j=1,\cdots, N,
\end{split}
 \end{eqnarray}
with $c_{j}$ and $d_{j}$ being arbitrary constants.
The solution of associated Lax pair system are given by
\begin{eqnarray}
\tilde{\phi}[N]=(\tilde{\phi}_{1}[N],\tilde{\phi}_{2}[N])^T=D[N]\tilde{\phi}[0],
\label{Nphi}
\end{eqnarray}
where
\begin{eqnarray}
\begin{split}
\tilde{\phi}[0]=\left(Ae^{-i\lambda x-2i\lambda^2t},Be^{i\lambda x+2i\lambda^2t}\right)^T,
\\
D[N]=\prod_{k=0}^{2N-1}\left((\lambda-\bar{\lambda}_{2N-k})I+(\bar{\lambda}_{2N-k}-\lambda_{2N-k})P[2N-k]\right),
\\
P[j]=\frac{1}{|\mu_j[j-1]|^2+|\nu_j[j-1]|^2}\left( \begin{array}{cc} |\mu_j[j-1]|^2 & \mu_j[j-1]\bar{\nu}_j[j-1] \\
 \bar{\mu}_j[j-1]\nu_j[j-1] & |\nu_j[j-1]|^2 \\ \end{array} \right),~j=1,\cdots,2N,
 \\
 \left( \begin{array}{c} \mu_j[j-1]  \\ \nu_j[j-1]  \\ \end{array} \right)
 =\prod_{k=1}^{j-1}\left.\left((\lambda-\bar{\lambda}_{j-k})I+(\bar{\lambda}_{j-k}-\lambda_{j-k})P[j-k]\right)\right|_{\lambda=\lambda_j}
 \left( \begin{array}{c} \mu_j \\ \nu_j  \end{array} \right),~j=1,\cdots,2N.
\end{split}
\end{eqnarray}
Inserting(\ref{ns}) and (\ref{Nphi}) into (\ref{DT}), we obtain the following $(N+1)$-soliton solution
\begin{eqnarray}
u=2i\frac{\Delta_2[N]}{\Delta_1[N]}-2i(\xi_1-\bar{\xi}_1)\left.\frac{\tilde{\phi}_{1}[N]\bar{\tilde{\phi}}_{2}[N]}{|\tilde{\phi}_{1}[N]|^2+|\tilde{\phi}_{2}[N]|^2}\right|_{\lambda=\xi_1},
\label{nstbc}
\end{eqnarray}
where $\xi_1=a+ib$, $\xi_1\neq \lambda_j$, $j=1,\cdots,2N$.
Such a solution satisfies the NLS equation together with the boundary condition (\ref{tbc}).

\section{Concluding remarks}

We have established the integrability of the NLS equation in the presence of time-dependent boundary condition (\ref{tbc}) both via the Sklyanin's formalism and via the tool of B\"{a}cklund transformations together with a reduction technique.
We also presented a direct method to construct soliton solutions of such a boundary problem. Our method for the construction of solutions is based on the Darboux transformation method in conjunction with a boundary dressing technique.
It is worth mentioning that an analogous time-dependent boundary for the integrable discrete NLS equation was studied very recently in \cite{CC2019},
where the nonlinear mirror image method \cite{BH2009,BB2012} was applied to construct the solutions.
We believe that the nonlinear mirror image method can also be applied to solve the boundary problem (\ref{tbc}) for the NLS equation.
The key point 
is to derive an additional symmetry (induced by the boundary) on the scattering data.
A possible way to derive such a symmetry is by using the fact that the  boundary (\ref{tbc}) can be interpreted as arising from
the B\"{a}cklund transformation connected two fields with an odd reduction (see section 2.2).
The study on this topic is not trivial, it will be performed in the future.

\section*{ACKNOWLEDGMENTS}
This work was supported by the National Natural Science Foundation of China (Grant No. 11771186).

\vspace{1cm}
\small{

}
\end{document}